\documentclass[aps,prd,twocolumn,groupedaddress]{revtex4}
\usepackage{graphicx}
\begin{document}

\title{Dirac Relaxation of the Israel Junction Conditions:\\
Unified Randall-Sundrum Brane Theory}

\author{Aharon Davidson}
\email[Email: ]{davidson@bgu.ac.il}
\author{Ilya Gurwich}
\email[Email: ]{gurwichphys@gmail.com}

\affiliation{Physics Department, Ben-Gurion University,
Beer-Sheva 84105, Israel}                 

\date{June 18, 2006}

\begin{abstract}
	Following Dirac's brane variation prescription, the
	brane must not be deformed during the variation
	process, or else the linearity of the variation may
	be lost.
	Alternatively, the variation of the brane is done, in a
	special Dirac frame, by varying the bulk coordinate
	system itself.
	Imposing appropriate Dirac style boundary conditions
	on the constrained 'sandwiched' gravitational action,
	we show how Israel junction conditions get relaxed, but
	remarkably, all solutions of the original Israel equations
	are still respected.
	The Israel junction conditions are traded, in the
	$Z_2$-symmetric case, for a generalized Regge-Teitelboim
	type equation (plus a local conservation law), and in the
	generic $Z_2$-asymmetric case, for a pair of coupled
	Regge-Teitelboim equations.
	The Randall-Sundrum model and its derivatives, such
	as the Dvali-Gabadadze-Porrati and the Collins-Holdom
	models, get generalized accordingly.
	Furthermore, Randall-Sundrum and Regge-Teitelboim
	brane theories appear now to be two different faces of
	the one and the same unified brane theory.
	Within the framework of unified brane cosmology, we
	examine the dark matter/energy interpretation of the
	effective energy/momentum deviations from General
	Relativity.
\end{abstract}

\maketitle   

\section{Introduction}
In an almost forgotten paper\cite{Dirac} entitled
\emph{''An extensible model of the electron''}, Dirac has
made an attempt to picture a classical spinless electron
as a breathing bubble in the electromagnetic field 'with
no constraints fixing its size and shape'.
Some positive surface tension has been invoked in order
'to prevent the electron from flying apart under the Coulomb
repulsion of it surface charge'.
On the practical side, this naive model has not made any impact
on particle physics, so in this paper, we make no use of the model
itself.
On the field theoretical side, however, although gravity was
switched off in this paper, it was nonetheless the first brane
model.
The equations of motion are derivable from an action principle,
and in particular, the model offers a detailed prescription how
to consistently perform brane variation.
In the present paper, we switch on gravity, and apply
the Dirac brane variation prescription to modern brane theories.

\begin{figure}[ht]
	\includegraphics[scale=1.35]{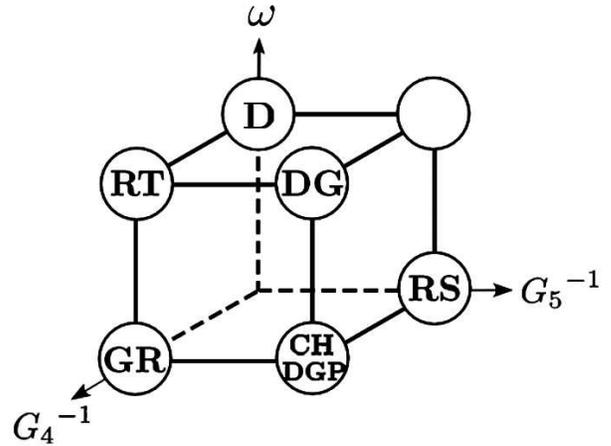}
	\caption{\label{cube} Randall-Sundrum (RS) and
	Regge-Teitelboim (RT) brane theories, which generalize
	General Relativity (GR), appear to be two different faces
	of the one and the same unified brane theory (DG).
	$\omega$ denotes a Regge-Teitelboim constant of integration.
	Also represented in the scheme are Dirac's extensible
	model (D), and the Dvali-Gabadadze-Porrati
	(DGP) and Collins-Holdom (CH) extensions of the RS model.}
\end{figure}

\smallskip
Our main result is that the Israel\cite{IJC} junction conditions
(IJC), which are known to play a central role in all modern brane
theories, get in fact relaxed.
While every IJC solution is still strictly respected, it represents
now a whole (continuous) family of new solutions.
This opens the door for brane unification.
To be more specific, as schematically illustrated in Fig.\ref{cube},
both Regge-Teitelboim\cite{RT} and Randall-Sundrum\cite{RS}
brane theories, which generalize General Relativity in very
different theoretical directions, appear to be two faces
of the one and the same unified brane theory.
Also represented in the scheme are Dirac model along with
Dvali-Gabadadze-Porrati\cite{DGP} and Collins-Holdom\cite{CH}
models.
These latter brane models switch on the Randall-Sundrum repressed
${\cal R}_{4}$ brane curvature term.

\smallskip
The Randall-Sundrum theory is very well known,
and has rightly attracted lots of attention from General
Relativity\cite{RSgr}, cosmology\cite{RScosmo} in particular,
and also from string theory\cite{RSstring} points of view.
See Ref.\cite{RSreview} for some brane world reviews,
and Ref.\cite{early} for some earlier brane models.
The much older Regge-Teitelboim theory, on the other hand,
a generalized Nambu-Goto type brane theory\cite{RT} for
quantum gravity, has remained quite unfamiliar.
This is partially due to the fact that the theory was originally
demonstrated within the naive framework of a flat (rather
than AdS) non-dynamical (rather than dynamical) higher
dimensional background.
The Regge-Teitelboim theory suffered some criticism\cite{Deser}
in the past, but given its built-in Einstein limit and a handful of
attractive features\cite{RTDavidson,RTothers} (in particular, a
dark companion\cite{RTDavidson} to any energy density, a
quadratic Hamiltonian formalism\cite{GBG}, and a rather
novel approach to brane nucleation\cite{nucleation}),
it offers a unique deviation from General Relativity.
The associated so-called Geodesic Brane Gravity (GBG)
is conveniently parametrized by a conserved 'bulk energy' (see
$\omega$ in Fig.1) which fades away at the Einstein limit.
Invoking the Dirac brane variation prescription, the appropriately
generalized Regge-Teitelboim theory naturally becomes the
$G_{5}\rightarrow\infty$ limit of the unified Randall-Sundrum
brane theory.

\smallskip
To see how all this comes about, let us first 'listen' very carefully
to Dirac.
Consider thus a flat $4$-dimensional Minkowski background
\begin{equation}
	ds^{2}=-dt^{2}+dr^{2}
	+r^{2}d\Omega^{2} ~,
\end{equation}
and denote the breathing radius of the Dirac bubble by
$r=f(t)$.
Alternatively, perform an implicit general coordinate transformation
$R=r-f(t)$, such that in the new frame, with metric
\begin{equation}
	ds^{2}=-dt^{2}+(dR+\dot{f}dt)^{2}
	+(R+f(t))^{2}d\Omega^{2} ~,
\end{equation}
the bubble location, conveniently set at $R=0$, does not change
during the variation process.
In both frames of reference the unknown function $f(t)$ is to be
determined by apparently minimizing the same Lagrangian, so
what is the difference?
According to Dirac, the trouble lies with the fact that the
inner ($R<0$) and the outer ($R>0$) regions of the bubble are
not a smooth continuation of each other.
In particular, originating from a close surface electric charge
distribution, the associated electric field solely lives outside the
bubble.
To be a bit more quantitative, introduce parameters $u_{1,2,3}$
to specify a general point on the bubble, and consider the
embedding vector $x^{\mu}(u)$ as a canonical variable.
The alarming point is that the variation $\delta I$ is not necessarily
a linear function of $\delta x^{\mu}$.
Quoting Dirac, \emph{\textbf{'If one makes a variation
$\delta x^{\mu}$ corresponding to the surface being pushed
out a little, $\delta I$ will not be minus the $\delta I$ for
$-\delta x^{\mu}$, corresponding to the surface being pushed
in a little, on account of the field just outside the surface being
different from the field just inside.
Thus this choice of canonical variables will not do'.}}
To bypass the problem, Dirac has ingeniously introduced
general curvilinear coordinates, such that in the new coordinate
system, the location of the bubble does not change during the
variation process.
The general idea being \emph{'to produce an arbitrary
variation of the surface by varying the coordinate system'.}
Modern brane theories based on an action principle better
follow this prescription or else the linearity, and hence the
self-consistency, of the variation may be in jeopardy.

\medskip
Although our paper is about relaxing the Israel junction
conditions, it is worth while mentioning the Battye-Carter
approach\cite{Carter}, where the brane junction conditions
are in fact further constrained.
In their model, the Israel junction conditions are not only
kept alive, but are furthermore supplemented by a force
equation.
The Battye-Carter force equation is trivially satisfied,
however, when $Z_{2}$ symmetry is enforced by hand.
In the present paper, for comparison, the Israel junction
conditions get relaxed, and the net effect does not die
away at the $Z_{2}$ limit.

\medskip
The introduction was mainly devoted to getting acquainted
with the Dirac brane variation.
The rest of the paper is organized as follows.
In Section 2, we covariantly formulate the Dirac brane
conditions.
In section 3, we support the Dirac brane prescription by
an explicit action principle (invoking a set of Lagrange
multipliers to manifestly deal with the various constraints
floating around), and carefully perform the variation without
deforming the brane, and without changing its location.
Next, Section 4 is where the Israel Junction Conditions
connect with the Regge-Teitelboim theory, and get
accordingly relaxed.
A detailed general analysis is offered without appealing to
a specific (say) $Z_2$ symmetry.
Unified brane cosmology is discussed in Section 5.
One integration of the equations of motion can be carried
out analytically, thereby introducing the novel constant of
integration $\omega$, which parametrizes the deviation
from the Randall-Sundrum model and of course from General
Relativity.
In Section 6, the various limits of the theory
(including in particular General Relativity, Randall-Sundrum,
and Regge-Teitelboim limits) are being discussed in some
details.
And finally, we summarize the paper by drawing some critical
conclusions, and sketching some exciting paths for future
research.

\section{Gravitational extension of
Dirac's boundary conditions}
Had we considered a curved, yet \emph{non-dynamical},
background (allow for different metrics $i=L,R$ on the
two sides of the brane), the discussion would have been very
much the same.
Following Dirac, given some fixed bulk metrics
$G^{i}_{AB}(y)$, the naive use of the embedding vectors
$y^{iA}(x)$ as canonical fields is physically inconsistent.
Alternatively, perform first the \emph{implicit} general
coordinate transformations $y^{iA}\rightarrow z^{ia}(y)$,
such that in the new Dirac $z$-frame, with bulk metrics
\begin{equation}
	G^{i}_{ab}(z)=
	G^{i}_{AB}(z(y))y^{iA}_{~,a}y^{iB}_{~,b} ~,
\end{equation}
the brane is kept at rest during the variation process.
The corresponding variations, namely
\begin{equation}
	\delta_{1}G^{ i}_{ab}=
	G^{ i}_{AB,C}\delta y^{iC}y^{iA}_{,a}y^{iB}_{,b}+
	2G^{ i}_{AB}y^{iA}_{,a}\delta y^{iB}_{,b} ~,
	\label{delta1}
\end{equation}
do not deform the brane.

\medskip
When bulk gravity is finally switched on, however, the
gravitational fields $G^{i}_{AB}$ themselves may vary (in addition
to the above variation associated with $\delta y^{iA}$), so there
are now \emph{two} different contributions to $\delta G_{ab}$,
namely
\begin{equation}
	\delta G_{ab}=
	\delta_{1} G_{ab}+\delta_{2}G_{ab}~,
\end{equation}
where
\begin{equation}
	\delta_{2}G_{ab} =
	y^{A}_{,a}y^{B}_{,b}\delta G_{AB} ~.
	\label{delta2}
\end{equation}
It is well known that, on the bulk, the arbitrariness of
$\delta G_{AB}$ leads to Einstein equations, whereas the
arbitrariness of $\delta y^A$, the essence of re-parametrization
covariance, is actually 'swallowed' in the sense that it does
not lead to any field equation.
But this by itself does not necessarily mean that the variation
must stay fully arbitrary on the brane as well.
In fact, we now argue that a fully arbitrary
$\delta G_{ab}{\big |}_{brane}$ would not only drive
the Dirac frame meaningless, but would furthermore violate
Dirac's 'linearity of the variation' principle.

\medskip
As long as $\delta G_{ab}{\big |}_{brane}$ is general enough and
is not restricted (to be more specific, it contains the full ten degrees
of freedom), the corresponding equation of motion, symbolically
written in the form
\begin{equation}
	\frac{\delta {\cal L}}
	{\delta G_{ab}}{\Big |}_{brane}= 0 ~,
	\label{symbolic}
\end{equation}
would strictly hold \emph{irrespective} of the
$\delta_{1}G_{ab}{\big |}_{brane}$ contribution.
This would practically close the door for performing
a consistent Dirac style $\delta y^{A}$ brane variation.
An elegant way out, and may be the only covariant way
out, is to impose the following (to be referred to as)
Dirac boundary conditions
\begin{equation}
	\fbox {$\delta_{2}G_{ab}{\Big |}_{brane}= 0 $}
	\label{Dirac}
\end{equation}
such that, on the brane, one would only tolerate the
arbitrariness of $\delta_{1}G_{ab}{\big |}_{brane}$.
The latter condition, which keeps the brane not deformed
during the variation process, is however not just a matter
of choice or convenience. It is in fact mandatory.
Rephrasing Dirac's argument, \emph{\textbf{if Eq.(\ref{Dirac})
is violated, a tiny deformation of the brane corresponding to
the brane being pushed a little to the right will not be minus
the variation corresponding to the brane being pushed a little
(equally) to the left, on account of the $L,R$ bulk sections not
being a smooth continuation of each other.}}
Eq.(\ref{Dirac}) paves thus the way for Dirac relaxation of Israel
junction conditions.
It should be noted that an earlier attempt\cite{Karasik} to
carry out brane variation Dirac style has unfortunately failed
to impose the boundary conditions Eq.(\ref{Dirac}), and
consequently the Israel junction conditions, although derived,
were not relaxed.

\medskip
A important remark is now in order.
In general, there are matter fields living in the bulk, and
obviously, their variation (in Dirac's $z$-frame) consists of two
parts.
For example, associated with some (say) vector field
\begin{equation}
	V_{a}(z)=V_{A}(z(y))y^{A}_{,a} ~,
\end{equation}
is the variation
\begin{equation}
	\delta V_{a}{\big |}_{bulk}=\delta V_{A}y^{A}_{,a}+
	V_{A,B}\delta y^{B}y^{A}_{,a}+V_{A}\delta y^{A}_{,a}~.
\end{equation}
Does it mean, in some (false) analogy with Eq.(\ref{Dirac}),
that one is obliged to further impose
\begin{equation}
	\delta_{2} V_{a}{\big |}_{brane}=
	y^{A}_{,a}\delta V_{A}{\big |}_{brane}=0 ~?
\end{equation}
Certainly not.
The gravitational field $G_{ab}$ is the only field in the theory
that, when being arbitrarily varied, it actually deforms the brane.
Once the brane is consistently put to rest during the variation
process, meaning Eq.(\ref{Dirac}) is fulfilled, Dirac's 'linearity of
the variation' principle is fully respected.
In turn, all matter fields may freely vary on the brane,
and thus, all conventional non-gravitational matching conditions
stay intact.

\section{Dirac-style brane variation}
Let the bulk line elements, on the two sides ($i=L,R$) of the
brane, be respectively
\begin{equation}
	ds^{2}_{i}=G^{i}_{AB}(y)dy^{iA}dy^{iB} ~.
\end{equation}
The brane metric gets then fixed once the two sets of embedding
coordinates $y^{iA}(x^{\mu})$ are consistently specified, so that
the common brane metric takes the form
\begin{equation}
	g_{\mu\nu}(x)
	= G^{L}_{AB}(y(x))y^{LA}_{~,\mu}y^{LB}_{~,\nu}
	= G^{R}_{AB}(y(x))y^{RA}_{~,\mu}y^{RB}_{~,\nu} ~.
\end{equation}
The fact that the brane metric is induced, rather than fundamental,
is crucial.
One may rightly conclude that, in a constrained brane gravity
theory, unlike in General Relativity, $g_{\mu\nu}(x)$ cannot
serve as a canonical field variable in the underlying Lagrangian
formalism.
Alternatively, the role of canonical fields is then naturally taken
by the embedding vectors $y^{iA}(x^{\mu})$, bearing in mind
that their physically consistent variation must be carried out Dirac
style.
Needless to say, if the higher dimensional background is furthermore
dynamical, the metric tensors $\displaystyle{G^{L,R}_{AB}(y)}$
enter the game as additional canonical fields.

\bigskip
The prototype brane action is the following
\begin{equation}
	\begin{array}{rcl}
	I & =  & \displaystyle{\int_{L}\left(
	-\frac{1}{16\pi G_{5}}{\cal R}^{L}+{\cal L}^{L}_{m}
	\right)\sqrt{-G_{L}}~d^{5}y
	~+} \vspace{6pt} \\
	& + &
	\displaystyle{\frac{1}{8\pi G_{5}}
	\int {\cal K}^{L}\sqrt{-g}~d^{4}x~+}
	\vspace{6pt} \\
	& +  & \displaystyle{\int\left(-\frac{1}{16\pi G_{4}}
	{\cal R}+{\cal L}_{m}
	\right)\sqrt{-g}~d^{4}x ~+}
	\vspace{6pt} \\
	& + & \displaystyle{\frac{1}{8\pi G_{5}}
	\int {\cal K}^{R}\sqrt{-g}~d^{4}x~+}
	\vspace{6pt} \\
	& + &  \displaystyle{\int_{R}\left(
	-\frac{1}{16\pi G_{5}}{\cal R}^{R}+{\cal L}^{R}_{m}
	\right)\sqrt{-G_{R}}~d^{5}y} 
	\end{array}
	\label{I}
\end{equation}
The constrained gravity brane Lagrangian is
'sandwiched' between the left and the right general
relativistic bulk Lagrangians (an early version of such
a 'sandwich' Lagrangian, describing a cosmic solenoid,
was considered in Ref.\cite{sandwich}).
The presence of the Gibbons-Hawking\cite{GH} boundary
terms here is known to be mandatory, not optional.
It is well known that without this exact term there is no
way to integrate out all $\delta G_{AB;C}$ terms on the
brane.
In our notations,  
\begin{equation}
	{\cal K}=g^{\mu\nu}{\cal K}_{\mu\nu}
	=P^{AB}{\cal K}_{AB}
\end{equation}
denotes the scalar extrinsic curvature,
\begin{equation}
	\begin{array}{c}
	\displaystyle{{\cal K}_{\mu\nu} =
	\frac{1}{2}y^{A}_{,\mu}y^{B}_{,\nu}
	(n_{A;B}+n_{B;A})=}   \vspace{4pt}\\
	\displaystyle{=\frac{1}{2}\left(
	y^{A}_{,\mu}n_{A,\nu}+y^{A}_{,\nu}n_{A,\mu}-
	2\Gamma^A_{BC}y^{B}_{,\mu}y^{C}_{,\nu}n_{A}
	\right)} 
	\end{array}
\end{equation}
is the extrinsic curvature tensor,
\begin{equation}
	P^{AB}=G^{AB}-n^{A}n^{B}=
	g^{\mu\nu}y^{A}_{,\mu}y^{B}_{,\nu}
\end{equation}
is the projection tensor (also known as the induced metric),
and $n^{A}$ being the pointing outwards space-like unit
normal to the brane.
The induced brane metric itself is given by both
\begin{equation}
	g_{\mu\nu}=G^L_{AB}y^{LA}_{~,\mu}y^{LB}_{~,\nu} 
	=G^R_{AB}y^{RA}_{~,\mu}y^{RB}_{~,\nu} ~.
	\label{g}
\end{equation}
Notice that in the Lagrangian specified by Eq.(\ref{I}), we keep
(i) The Randall-Sundrum option of throwing away
the 4-dimensional Ricci scalar ${\cal R}$ (by letting
$G_{4}\rightarrow\infty$), and
(ii) The Regge-Teitelboim option of turning the
higher dimensional background non-dynamical (by letting
on $G_{5}\rightarrow\infty$).

\medskip
The canonical fields associated with the above action, are
$G^{L,R}_{AB}(y)$, and the brane location $y^{A}(x)$ (to be
treated as prescribed by Dirac).
At the technical level, however, given the various constrains
floating around, e.g. Eq.(\ref{g}), we find this Lagrangian
too complicated to handle.
Thus, to make the constraints manifest, we invoke a set of
Lagrange multipliers, and correspondingly add an extra piece
$I_{con}$ to the action, namely
\begin{equation}
	\fbox{$
	\begin{array}{c}
	I_{con}  =   \displaystyle{\sum_{i=L,R}\bigg[
	\int \lambda^{\mu\nu}_i \left(
	g_{\mu\nu}-G^i_{AB}y^{iA}_{~,\mu}y^{iB}_{~,\nu}\right)~+}
	\vspace{2pt} \\
	+~\displaystyle{\eta^{\mu}_{i}y^{iA}_{~,\mu}n^{i}_{A}
	+\sigma_{i}\left(G^{i}_{AB}n^{iA}n^{iB}-1\right)\bigg]
	\sqrt{-g}~d^{4}x} 
	\end{array}$}
	\label{DeltaI}
\end{equation}
In this alternative formalism, $\lambda^{\mu\nu}_{L,R}$,
$\eta^{\mu}_{L,R}$, and $\sigma_{L,R}$ denote Lagrange
multipliers, whereas $G^{L,R}_{AB}(y)$, $n^{L,R}_{A}(y)$,
and $g_{\mu\nu}(x)$ are considered \emph{independent}
canonical fields.
In particular, it is crucial to emphasize that the embedding
vectors $y^A_{L}(x)$ and $y^A_{R}(x)$ (like $n^A_{L}$
and $n^A_{R}$) are kept independent at the Lagrangian
level, leading to two field equations.
It is only when the constrains associated with the Lagrange
multipliers $\lambda_{\mu\nu}^{L,R}$ are written down
explicitly, at the equation of motion level, that the two metrics
$g^{L,R}_{\mu\nu}(x)$, induced by $y^A_{L}(x)$ and
$y^A_{R}(x)$ respectively, become in fact the one and the
same brane metric $g_{\mu\nu}(x)$.

\medskip
We now perform the variation in steps:

\medskip
$\bullet$ The variation of $I+I_{con}$ with respect to the
normals $n^{L,R}_{A}$ leads to the field equation (for $L,R$
separately)
\begin{equation}
	-\frac{1}{8\pi G_{5}}g^{\mu\nu}\left(y^{A}_{;\mu\nu}+
	\Gamma^{A}_{BC}y^{B}_{,\mu}y^{C}_{,\nu}\right)+
	\eta^{\mu}y^{A}_{,\mu}+2\sigma n^{A}=0 ~.
\end{equation}
To squeeze its contents, one may project this equation on
the brane (i.e. multiplying by $G_{DA}y^{D}_{,\lambda}$) to
learn that
\begin{equation}
	\eta_{\mu}=0 ~,
\end{equation}
and also perpendicular to the brane (i.e. multiplying by
$n_{A}$) to find out that
\begin{equation}
	\sigma=-\frac{1}{16\pi G_{5}}{\cal K} ~.
	\label{sigma}
\end{equation}

$\bullet$ The variation of $I+I_{con}$ with respect to the
brane metric $g_{\mu\nu}$ contains contributions from both
$i=L,R$, and after taking the various constraints into account,
one is left with the field equation
\begin{equation}
	\begin{array}{c}
	\displaystyle{\frac{1}{16\pi G_{4}}\left(
	{\cal R}^{\mu\nu}-\frac{1}{2}g^{\mu\nu}{\cal R}\right)
	+\frac{1}{2}{\cal T}^{\mu\nu}+}
	\vspace{4pt}\\
	\displaystyle{+~\sum_{i=L,R}\lambda^{\mu\nu}_{i}-
	\frac{1}{8\pi G_{5}}\sum_{i=L,R}\left(
	{\cal K}^{\mu\nu}_{i}-
	\frac{1}{2}g^{\mu\nu}{\cal K}_{i}\right)=0 ~.}
	\end{array}
	\label{gvariation}
\end{equation}

$\bullet$ The bulk variation of $I+I_{con}$ with respect to
the metrics $G^{L,R}_{ab}$ is next.
It is clear from our notations that, at this stage, it is already
mandatory to follow the Dirac prescription.
A useful formula here is the contraction of a general tensor
$E^{ab}$ with the variation of the metric $\delta G_{ab}$
\begin{equation}
	\begin{array}{c}
	E^{ab}\delta G_{ab} = E^{AB}\delta G_{AB} -
	\vspace{6pt} \\
	 -2(E^{AB})_{;B}G_{AD}\delta y^{D} 
	+2\left(E^{ab}G_{AB}y^{A}_{\,,a}\delta y^{B}\right)_{\,;b} ~.
	\end{array}
	\label{E}
\end{equation}
In particular, applying the above to 
\begin{equation}
	E_{i}^{ab}=\frac{1}{8\pi G_{5}}\left(
	{\cal R}_{i}^{ab}-\frac{1}{2}G_{i}^{ab}{\cal R}_{i}\right)
	+{\cal T}_{i}^{ab} ~,
\end{equation}
we learn that there are no surprises in the two $i=L,R$
bulk sections, which are conventionally governed by the
5-dimensional Einstein equations $E_{i}^{AB}=0$.
Notice that the total derivative term in Eq.(\ref{E}) can be
transformed into a boundary term by means of Stokes's 
theorem, and thus may in principle contribute to the
$\delta G_{ab}$ variation on the brane.
But such a contribution clearly vanishes due to the fact that
on the bulk Einstein equations are satisfied. 

\medskip
$\bullet$ The variation of $I+I_{con}$ with respect to the
metrics $G^{L,R}_{ab}{\big|}_{brane}$ (for $L,R$ separately)
is where Dirac boundary conditions Eq.(\ref{Dirac}) are
expected to make their impact.
Given the above action principle, one can now explicitly
specify the corresponding integrant, namely
\begin{equation}	
	\begin{array}{rcl}
	\displaystyle{\frac{1}{\sqrt{-g}}
	\frac{\delta {\cal L}}{\delta G_{ab}}{\Big|}_{brane}}&=&       
	\displaystyle{\frac{1}{16\pi G_{5}}\left[
	n^{a}g^{\mu\nu}\left(y^{b}_{;\mu\nu}+
	\Gamma^{b}_{cd}y^{c}_{,\mu}y^{d}_{,\nu}
	\right)+\right.}
	\vspace{6pt} \\
	&+&\displaystyle{\left. n^{a}_{;c}P^{bc}\right]-
	\lambda^{\mu\nu}y^{a}_{,\mu}y^{b}_{,\nu}
	-\sigma n^{a}n^{b}  ~.}
	\end{array}
	\label{branevariation}
\end{equation}
Had we equated the RHS of Eq.(\ref{branevariation})
to zero, which is however \emph{not} the case here, the
normal-normal component
$\displaystyle{\frac{\delta {\cal L}}{\delta G_{ab}}n_{a}n_{b}}$
would
have re-produced Eq.(\ref{sigma}), the normal-tangent
components $\displaystyle{\frac{\delta {\cal L}}{\delta G_{ab}}
G_{ac}y^{c}_{,\alpha}n_{b}}$ would have vanished identically,
and the tangent-tangent components
$\displaystyle{\frac{\delta {\cal L}}
{\delta G_{ab}}G_{ac}y^{c}_{,\alpha}G_{bd}y^{d}_{,\beta}}$
would have resulted in
\begin{equation}
	\frac{1}{16\pi G_{5}}{\cal K}_{\mu\nu}
	-\lambda_{\mu\nu}=0 ~,
	\label{lambdaold}
\end{equation}
thereby taking us directly, see Eq.(\ref{gvariation}), to Israel
junction conditions.

\medskip
At this point, we find it quite amazing to recall that the Israel
junction conditions\cite{IJC} were originally introduced without
the support of an underlying action principle.
For some early derivation of IJC from a gravitational action
principle, see for example Ref.\cite{sandwich}, and
Ref.\cite{dsandwich} for a dilatonic version and other extensions.

\medskip
However, as argued earlier, insisting on Dirac boundary
conditions Eq.(\ref{Dirac}), the RHS of Eq.(\ref{branevariation})
does not necessarily vanish.
To derive the modified field equation, we first substitute
$\sigma$, as given by Eq.(\ref{sigma}), so that the variation
on the brane takes a more compact form
\begin{equation}
	\int \sqrt{-g}\left(\frac{1}{16\pi G_{5}}{\cal K}^{ab}
	-\lambda^{\mu\nu}y^{a}_{,\mu} y^{b}_{,\nu}\right)
	\delta_1 G_{ab}{\Big |}_{brane}d^{4}x=0 ~.
\end{equation}
We know that now, once $\sigma$ has been already substituted,
the normal-normal component of the tensor in parentheses vanishes,
while its normal-tangent components vanish identically.
Thus, the only contribution to the integral comes from the
tangent-tangent components, namely
\begin{equation}
	\int \sqrt{-g}\left(\frac{1}{16\pi G_{5}}{\cal K}^{\mu\nu}
	-\lambda^{\mu\nu}\right)
	\delta_1 G_{ab}{\Big |}_{brane}y^{a}_{,\mu} y^{b}_{,\nu}~
	d^{4}x=0 ~.
\end{equation}
Expressing $\delta_1 G_{ab}{\big |}_{brane}$ in terms of
$\delta y^{A}$ and its derivatives, see Eq.(\ref{delta1}),
we are finally led to
\begin{equation}
	\begin{array}{l}
  	\displaystyle{\int \sqrt{-g}
	\left(\frac{1}{16\pi G_{5}}{\cal K}^{\mu\nu}
	-\lambda^{\mu\nu}\right)} \vspace{6pt}\\
  	\displaystyle{\quad\left(G_{AB,C}y^{A}_{,\mu}
	y^{B}_{,\nu}\delta y^{C}+
	2G_{AB}y^{A}_{,\mu}\delta y^{B}_{,\nu}
	\right)d^{4}x=0} ~.
	\end{array}
\end{equation}
At this stage, no trace is left from the Dirac frame (that is to
show that it is no more than a mathematical tool), and the
variation acquires the familiar Regge-Teitelboim form.
It is now straight forwards to integrate by parts, and verify that
the arbitrariness of $\delta y^{A}$ gives rise to
\begin{equation}
	\begin{array}{c}
	\displaystyle{\left[\left(\frac{1}{16\pi G_{5}}{\cal K}^{\mu\nu}
	-\lambda^{\mu\nu}\right)
	y^{A}_{,\mu}\right]_{;\nu}+ } \vspace{4pt} \\
	\displaystyle{+~\Gamma^{A}_{BC}
	\left(\frac{1}{16\pi G_{5}}{\cal K}^{\mu\nu}
	-\lambda^{\mu\nu}\right)
	y^{B}_{,\mu}y^{C}_{,\nu}=0 ~,}  
	\end{array}
	\label{va}
\end{equation}
for $i=L,R$ separately.
Reflecting a fundamental embedding identity, the
velocity $y^{A}_{,\mu}$ and the covariant acceleration
$y^{A}_{;\mu\nu}+\Gamma^{A}_{BC}y^{B}_{,\mu}y^{C}_{,\nu}$,
viewed as vectors on the bulk, are orthogonal to each other.
In turn, Eq.(\ref{va}) splits into two parts, each of which must
vanish separately.
Whereas the part proportional to $y^{A}_{,\mu}$ implies the local
conservation law
\begin{equation}
	\left(\frac{1}{16\pi G_{5}}{\cal K}^{\mu\nu}
	-\lambda^{\mu\nu}\right)_{;\nu}=0 ~,
	\label{conserve}
\end{equation}
the second part is the geodesic brane equation
\begin{equation}
	\left(\frac{1}{16\pi G_{5}}{\cal K}^{\mu\nu}
	-\lambda^{\mu\nu}\right)
	\left(y^{A}_{;\mu \nu}+
	\Gamma^{A}_{BC}y^{B}_{,\mu}y^{C}_{,\nu}\right)=0 ~.
\end{equation}
Appreciating the fact that the extrinsic curvature is the normal
component of the covariant acceleration, the latter equation
takes the more geometric oriented form
\begin{equation}
	\left(\frac{1}{16\pi G_{5}}{\cal K}_{\mu\nu}
	-\lambda_{\mu\nu}\right){\cal K}^{\mu\nu}=0 ~.
	\label{lambdanew}
\end{equation}
Altogether, the tensorial Eq.(\ref{lambdaold}), associated
with Israel junction conditions, is replaced by the
${\cal K}$-contracted scalar Eq.(\ref{lambdanew}) plus the
conservation law Eq.(\ref{conserve}).

\section{Relaxing the Israel junction conditions}
Define the combined Einstein-Israel tensor
\begin{equation}
	\begin{array}{c}
	\displaystyle{ {\cal S}_{\mu\nu} \equiv
	\frac{1}{16\pi G_{4}}
	\left({\cal R_{\mu\nu}}
	-\frac{1}{2}g_{\mu\nu}{\cal R}\right)}+
	\vspace{4pt} \\
	\displaystyle{+\frac{1}{2}{\cal T}_{\mu\nu}
	-\frac{1}{16\pi G_{{5}}}\sum_{i=L,R}
	\left({\cal K}^{i}_{\mu\nu}
	-{\cal K}^{i}g_{\mu\nu}\right)}~.
	\end{array}
	\label{S}
\end{equation}
In this convenient notation, our gravitational field equations
take the compact form	
\begin{equation}
	\fbox{$\displaystyle{{\cal S}_{\mu\nu}=\sum_{i=L,R}
	\left(\frac{1}{16\pi G_{{5}}}{\cal K}^{i}_{\mu\nu}
	-\lambda^{i}_{\mu\nu}\right) }$}
	\label{master}
\end{equation}
Now, if Dirac's brane variation procedure is ignored, then
the RHS vanishes, and one recovers the well known Israel
junction conditions ${\cal S}_{\mu\nu}=0$.
However, if the Dirac brane variation procedure is adopted,
the RHS does not necessarily vanish, and one is alternatively
led to
\begin{eqnarray}
	\left(\frac{1}{16\pi G_{{5}}}
	{\cal K}^{L}_{\mu\nu}-\lambda^{L}_{\mu\nu}\right)
	{\cal K}_{L}^{\mu\nu}=0 ~,
	\label{lambda1a}\\
	\left(\frac{1}{16\pi G_{{5}}}
	{\cal K}^{R}_{\mu\nu}-\lambda^{R}_{\mu\nu}\right)
	{\cal K}_{R}^{\mu\nu}=0 ~,
	\label{lambda1b}
\end{eqnarray}
accompanied by two independent conservation laws
\begin{eqnarray}
	\left(\frac{1}{16\pi G_{{5}}}{\cal K}_{L}^{\mu\nu}
	-\lambda_{L}^{\mu\nu}\right)_{;\nu}=0 ~,
	\label{lambda2a}\\
	\left(\frac{1}{16\pi G_{{5}}}{\cal K}_{R}^{\mu\nu}
	-\lambda_{R}^{\mu\nu}\right)_{;\nu}=0 ~.
	\label{lambda2b}
\end{eqnarray}
Notice that Eqs.(\ref{lambda1a},\ref{lambda1b}) cause the
master Eq.(\ref{master}) to split into two projections, namely
\begin{equation}
	\fbox{$\begin{array}{c}
	\displaystyle{{\cal S}_{\mu\nu}{\cal K}_{L}^{\mu\nu}
	= \left(\frac{1}{16\pi G_{5}}{\cal K}^{R}_{\mu\nu}
	-\lambda^{R}_{\mu\nu}\right)
	{\cal K}_{L}^{\mu\nu}}  \vspace{4pt}\\
	\displaystyle{{\cal S}_{\mu\nu}{\cal K}_{R}^{\mu\nu}
	= \left(\frac{1}{16\pi G_{5}}{\cal K}^{L}_{\mu\nu}
	-\lambda^{L}_{\mu\nu}\right)
	{\cal K}_{R}^{\mu\nu}}
	\end{array}$} 
	\label{SLR}
\end{equation}
whereas the major role of Eqs.(\ref{lambda2a},\ref{lambda2b}),
combined with the Coddazi relations, see Eq.(\ref{Coddazi}), is
to assure generalized energy/momentum conservation even
though Einstein and Israel equations get relaxed.

\bigskip
Given the set of Eqs.(\ref{lambda1a}-\ref{SLR}), it is in general
impossible to construct a generic brane equation free of the
Lagrange multipliers $\lambda^{L,R}_{\mu\nu}$.
However, there are two special cases for which the calculation
of $\lambda^{L,R}_{\mu\nu}$ can be practically bypassed.
These special cases are the following:

\medskip
\noindent
\underline{The smooth background case:}

\medskip
This 'mostly pedagogical' case is characterized by
\begin{equation}
	{\cal K}^{L}_{\mu\nu}+{\cal K}^{R}_{\mu\nu}=0~,
\end{equation}
so that
\begin{equation}
	\left(\lambda^{L}_{\mu\nu}+
	\lambda^{R}_{\mu\nu}\right)
	{\cal K}^{\mu\nu}=0 ~.
\end{equation}
Associated with this case is the original (to be referred to
as the reduced) Regge-Teitelboim equation written in a
geometrically oriented form\cite{Carter}
\begin{equation}
	\left({\cal R_{\mu\nu}}-\frac{1}{2}g_{\mu\nu}{\cal R}
	+8\pi G_{4}{\cal T}_{\mu\nu}\right)
	{\cal K}^{\mu\nu}=0 ~.
	\label{reduced}
\end{equation}
The Einstein limit, a crucial built-in feature of the Regge-Teitelboim
theory, is manifest (for a finite $G_{4}$), with the Newton
constant being easily identified as
\begin{equation}
	G_N=G_4 ~.
\end{equation}
In particular, it is evident that every solution of Einstein equations
is automatically a solution of the corresponding (reduced)
Regge-Teitelboim equation.
The deviation from General Relativity is expected to be parametrized
by a novel constant of integration (details soon) .

\medskip
\noindent
\underline{The $Z_{2}$-symmetric case:}

\medskip
This 'more practical' case is characterized by
\begin{equation}
	{\cal K}^{L}_{{\mu\nu}}-{\cal K}^{R}_{\mu\nu}=0~,
\end{equation}
for which
\begin{equation}
	\left(\lambda^{L}_{\mu\nu}-
	\lambda^{R}_{\mu\nu}\right)
	{\cal K}^{\mu\nu}=0 ~.
\end{equation}
Associated with this case is the generalized (to be referred
to as the full) Regge-Teitelboim equation
\begin{equation}
	{\cal S}_{\mu\nu}{\cal K}^{\mu\nu}=0 ~,
	\label{full}
\end{equation}
where the Einstein-Israel tensor ${\cal S}_{\mu\nu}$ has been
defined by Eq.(\ref{S}).
As expected, it manifestly exhibits a Randall-Sundrum limit,
with or without the Einstein tensor contribution (which anyhow
dies away as $G_4 \rightarrow\infty$).
In this limit, Following Collins-Holdom \cite{CH}, one may
rightly expect (to be re-derived later) the Newton constant to
obey
\begin{equation}
	\frac{1}{G_{N}}=\frac{1}{G_{RS}}+\frac{1}{G_{4}}~,
	\label{GN}
\end{equation}
modifying its original Randall-Sundrum value
\begin{equation}
	G_{RS}=G_{5}\sqrt{-\frac{\Lambda_{5}}{6}} ~.
\end{equation}
Clearly, every solution of Israel junction conditions is now
automatically
a solution of the corresponding (full) Regge-Teitelboim equation.
Also notice that we meet again the reduced Regge-Teitelboim
equation at the limit where $G_5 \rightarrow\infty$.

\medskip
On pedagogical grounds, we find it convenient to define a asterisked
energy/momentum tensor
\begin{equation}
	{\cal T}^{\displaystyle \ast}_{\mu\nu} \equiv {\cal T}_{\mu\nu}
	-\frac{1}{8\pi G_{{5}}}\sum_{i=L,R}
	\left({\cal K}^{i}_{\mu\nu}-{\cal K}^{i}g_{\mu\nu}\right) ~,
	\label{star}
\end{equation}
and asterisked Lagrange multipliers
\begin{equation}
	\lambda^{i\displaystyle \ast}_{\mu\nu} \equiv
	\lambda^{i}_{\mu\nu}
	-\frac{1}{8\pi G_{{5}}}
	{\cal K}^{i}_{\mu\nu} ~.
\end{equation}
In this language, ${\cal T}^{\displaystyle \ast}_{\mu\nu}=0$ is
recognized as the main brane equation in the original Randall-Sundrum
model, where intrinsic brane gravity has been switched off
($G_{4}\rightarrow\infty$).
${\cal T}^{\displaystyle \ast}_{\mu\nu}$ resumes a more
natural role as an implicit energy/momentum tensor in the
dressed (finite $G_{4}$) Randall-Sundrum model.
Invoking the above definition, the full Regge-Teitelboim
equation resembles its reduced (albeit asterisked) version which,
for the $Z_{2}$ symmetric case of interest, takes the form
\begin{equation}
	\left({\cal R_{\mu\nu}}-\frac{1}{2}g_{\mu\nu}{\cal R}
	+8\pi G_{4}{\cal T}^{\displaystyle \ast}_{\mu\nu}\right)
	{\cal K}^{\mu\nu}=0 ~.
	\label{starRT}
\end{equation}
Stemming from $\lambda^{{\displaystyle \ast}i}_{~\mu\nu}$
conservation, the integrability condition of the resulting
Regge-Teitelboim equation requires
${\cal T}^{\displaystyle \ast}_{\mu\nu}$ (not necessarily
${\cal T}_{\mu\nu}$) to be locally conserved on the brane
\begin{equation}
	{\cal T }^{{\displaystyle \ast}\mu\nu}_{~~~;\nu}=0 ~.
	\label{starcon}
\end{equation}
This is guaranteed in fact  by Eqs.(\ref{lambda2a},\ref{lambda2b}),
or more precisely by the Coddazi relation
\begin{equation}
	\begin{array}{rcl}
	{\cal T }^{\mu\nu}_{\quad;\nu} &=&
	\displaystyle{\frac{1}{8\pi G_{{5}}}\sum_{i=L,R}
	\left({\cal K}^{\mu\nu}_{i}
	-{\cal K}_{i}g^{\mu\nu}\right)_{;\nu}}~= 
	\vspace{4pt}\\
	&=& \displaystyle{\frac{1}{8\pi G_{{5}}}g^{\mu\nu}
	\sum_{i=L,R}n^{a}_{i}{\cal R}^{i}_{ab}y^{ib}_{~,\nu}}~.
	\end{array}
	\label{Coddazi}
\end{equation}
Given the special case of a maximally symmetric embedding
spacetime (such as AdS), ${\cal T}_{\mu\nu}$ itself happens
to be conserved.

\section{Unified brane cosmology}
In this section we deal with brane cosmology which is governed
by a generic perfect fluid energy/momentum tensor.
As usual, the latter is characterized by some energy density
$\rho$, and isotropic pressure $P$.
No specific equation of state $P=P(\rho)$ is chosen at this stage.
The cosmological brane metric, which takes the standard FRW form
\begin{equation}
	ds^{2}=-dt^{2}+
	\frac{a^{2}(t)}{\left(1+\frac{1}{4}kr^{2}\right)^{2}}
	\delta_{ij} dx^{i}dx^{j} ~,
\end{equation}
is embedded within a $Z_{2}$ symmetric AdS background with a
negative cosmological constant $\Lambda_{5}<0$.
The associated extrinsic curvatures are given explicitly by
\begin{equation}
	{\cal K}^{L,R}_{\mu\nu}=\left(
	\begin{array}{cc}
	\displaystyle{\frac{1}{\xi}\left(\frac{\ddot{a}}{a}-
	\frac{\Lambda_{5}}{6}\right)}  & 0    \\
	0 & \displaystyle{-\frac{a^{2}\xi}
	{\left(1+\frac{1}{4}kr^{2}\right)^{2}}\delta_{ij}}    
	\end{array}
	\right) ~,
\end{equation}
where we have used the convenient notation
\begin{equation}
	\xi \equiv \sqrt{\frac{\dot{a}^{2}+k}{a^{2}}-
	\frac{\Lambda_{5}}{6}} ~.
	\label{xi}
\end{equation}

\medskip
Given the pair $(\rho,P)$ and the above expressions for
${\cal K}^{L,R}_{\mu\nu}$, we thus follow Eq.(\ref{star}), and
define the asterisked pair
$(\rho^{\displaystyle\ast},P^{\displaystyle\ast})$ via
\begin{equation}
	\begin{array}{rcl}
	\rho^{\displaystyle\ast} & = & \displaystyle{\rho-
	\frac{3\xi}{4\pi G_{5}} } ~,  \\
	P^{\displaystyle\ast} & = & \displaystyle{P+\frac{1}{4\pi G_{5}}
	\left(2\xi+\frac{1}{\xi}
	\left(\frac{\ddot{a}}{a}-\frac{\Lambda_{5}}{6}\right)\right)} ~.
	\end{array}
	\label{stardef}
\end{equation}
We have already proven, see Eq.(\ref{starcon}), the validity of the
local conservation law
\begin{equation}
	\dot{\rho}^{\displaystyle\ast}+
	3\frac{\dot{a}}{a}\left(\rho^{\displaystyle\ast}+
	P^{\displaystyle\ast}\right)=0 ~,
\end{equation}
which, owing to the maximally symmetric geometry of both bulk
sections, is equivalent to
\begin{equation}
	\dot{\rho}+3\frac{\dot{a}}{a}\left(\rho+P\right)=0 ~.
\end{equation}

\medskip
The main equation for the FRW scale factor $a(t)$, namely
Eq.(\ref{starRT}), can now be put together, and one finds
\begin{equation}
	\begin{array}{l}
	\displaystyle{\left(8\pi G_{4}\rho^{\displaystyle\ast}-
	3\frac{\dot{a}^{2}+k}{a^{2}}\right){\cal K}_{tt}~+} \vspace{6pt}\\
	\displaystyle{+~\frac{\left(1+\frac{1}{4}kr^{2}\right)^{2}}{a^{2}}
	\left(8\pi G_{4}P^{\displaystyle\ast}+2\frac{\ddot{a}}{a}+
	\frac{\dot{a}^{2}+k}{a^{2}}
	\right)\delta^{ij}{\cal K}_{ij}=0} ~.
	\end{array}
\end{equation}
After some algebra, and ${\cal K}_{\mu\nu}$ substitution, it can be
rearranged into
\begin{equation}
	\begin{array}{l}
	  \displaystyle{{\Big (}8\pi G_{4}\rho^{\displaystyle\ast}a^{2}-
	  3(\dot{a}^{2}+k){\Big )}
	\left(\ddot{a}a-\frac{\Lambda_{5}}{6}a^{2}\right)-}
	\vspace{4pt} \\
	-  \displaystyle{3{\Big (}8\pi G_{4}P^{\displaystyle\ast}a^{2}+
	2\ddot{a}a+(\dot{a}^{2}+k){\Big )}
	\left(\dot{a}^{2}+k-\frac{\Lambda_{5}}{6}a^{2}\right) =0 ~.}
	\end{array}
\end{equation}
We have thus encountered a second order differential equation
which we would now like to integrate out.
To be more precise, recalling ${\cal T}^{\displaystyle\ast}_{\mu\nu}$
conservation and the fact that
$\rho^{\displaystyle\ast}=\rho^{\displaystyle\ast}(\dot{a},a)$,
we are after some function
\begin{equation}
	F(\rho^{\displaystyle\ast},\dot{a},a)=const.
\end{equation}
The analytic answer\cite{RTDavidson} can be borrowed from
the Regge-Teitelboim theory
(only with $\rho^{\displaystyle\ast}$ replacing now $\rho$),
namely
\begin{equation}
	F=a^{4}\left(
	3\frac{\dot{a}^{2}+k}{a^{2}}-
	8\pi G_{4}\rho^{\displaystyle\ast}\right)
	\sqrt{\frac{\dot{a}^{2}+k}{a^{2}}-
	\frac{\Lambda_{5}}{6}},
\end{equation}
and the same holds for the rest of the analysis.
Consequently, it is very useful to define
$\rho^{\displaystyle\ast}_{d}$ (to be referred to as the 'dark'
companion of $\rho^{\displaystyle\ast}$) by means of
\begin{equation}
	\dot{a}^{2}+k \equiv \frac{8\pi G_{4}}{3}
	\left({\rho^{\displaystyle\ast}}+
	{\rho^{\displaystyle\ast}_{d}}\right)a^{2} ~.
	\label{starFRW}
\end{equation}
The justification for using here the word 'dark' is to be discussed
soon.
Plugging now the latter definition into
$F(\rho^{\displaystyle\ast},\dot{a},a)$, one is immediately led
to a the result
\begin{equation}
	\fbox{$\displaystyle{
	{\rho^{\displaystyle\ast}_{d}}^{2}
	\left(8\pi G_{4}\left({\rho^{\displaystyle\ast}}+
	{\rho^{\displaystyle\ast}_{d}}\right)-
	\frac{\Lambda_{5}}{2}\right)=
	\frac{\omega^{2}}{a^{8}} }$} 
	\label{omega}
\end{equation}
where the novel constant of integration $\omega$ serves to
parametrize the deviations from Randall-Sundrum brane cosmology.

\medskip
Regarding the physical interpretation, imagine a physicist equipped
with Einstein gravitational field equations, but totally ignorant of the
existence of a fifth dimension and the associated brane gravity.
Our physicist may have already calculated the Newton constant
$G_{N}$, and is presumably capable of measuring the observed
energy density $\rho(a)$.
Verifying Einstein equations experimentally, that is establishing
the ultimate interplay between geometry and matter/energy, is all
what our physicist hopes for.
Failing to do so consistently, however, with a discrepancy level of
$90\%$, he would presumably refer to the missing ingredient of
General Relativity as $\rho_{dark}$, the so-defined illusive dark
companion of observable $\rho$, thereby enforcing an effective
FRW evolution based on
\begin{equation}
	\dot{a}^{2}+k \equiv \frac{8\pi G_{N}}{3}
	\left({\rho}+{\rho_{dark}}\right)a^{2} ~.
	\label{darkFRW}
\end{equation}
It remains to be seen if unified brane gravity has anything
to do with reality, but until then, one cannot resist the
speculation that the missing dark matter is nothing but
a brane artifact (various alternatives to dark matter and dark
energy are reviewed in Ref.\cite{dark}).
In other words, subject to the correct identification of $G_{N}$,
unified brane gravity suggests
\begin{equation}
	\rho_{dark}= \frac{G_{4}}{G_{N}}(\rho^{\displaystyle\ast}+
	\rho^{\displaystyle\ast}_{d})-\rho ~.
\end{equation}
Note that the idea of dark matter/energy unification\cite{unified},
that is the possibility that artifact dark matter is nothing but the
dark companion of $\rho=\Lambda_{4}$, been discussed within
the framework of the original Regge-Teitelboim model.

\medskip
Finally, we derive the cubic equation
\begin{equation}
	\fbox{$\displaystyle{\frac{3\xi^{3}}{8\pi G_{4}}+
	\frac{3\xi^{2}}{4\pi G_{5}}+
	\left(\frac{\Lambda_{5}}{16\pi G_{4}}-\rho(a)\right)\xi+
	\frac{\omega}{\sqrt{3}a^{4}}=0 }$}
	\label{cubic}
\end{equation}
which directly relates $\xi$, as defined by Eq.(\ref{xi}), to the
bare energy density $\rho(a)$.
In turn, the FRW equation takes the effective form
\begin{equation}
	\frac{\dot{a}^{2}+k}{a^{2}}=
	\frac{\Lambda_{5}}{6}+\xi^2 (a) ~.
\end{equation}
On the practical side, Eq.(\ref{cubic}) allows us to conveniently
navigate within the $\left( G_{4}^{-1},G_{5}^{-1},\omega \right)$
parameter space, with all special cases easily accessible.

\section{GR, RT, RS, DGP, and CH limits}
\subsection{Maximally symmetric brane}
To get a glimpse of what is lying ahead, let us first study the
pedagogical case where the FRW cosmological brane evolution
is enforced to be solely governed by a cosmological constant,
namely
\begin{equation}
	\dot{a}^2+k=\frac{1}{3}\Lambda_{4}a^2 ~.
\end{equation}
In which case, Eq.(\ref{starFRW}) immediately tells us that
\begin{equation}
 	8\pi G_{4}\left(\rho^{\displaystyle\ast}+
	\rho^{\displaystyle\ast}_{d} \right)=\Lambda_{4} ~,
\end{equation}
which can be further translated, as dictated by Eq.(\ref{omega}),
into
\begin{equation}
	\begin{array}{l}
	\displaystyle{\rho^{\displaystyle\ast}=
	\frac{\Lambda_{4}}{8\pi G_{4}}+
	\frac{\omega}{a^{4}\sqrt{\Lambda_{4}-
	\frac{1}{2}\Lambda_{5}}}}~,
	\vspace{4pt} \\
	\displaystyle{\rho^{\displaystyle\ast}_{d} =
	-\frac{\omega}{a^{4}\sqrt{\Lambda_{4}-
	\frac{1}{2}\Lambda_{5}}}}
	\end{array}
\end{equation}
This simply means, following Eq.(\ref{stardef}), that one
should have actually started from the primitive energy density
\begin{equation}
	\rho (a)=\frac{\Lambda_{4}}{8\pi G_{4}}+
	\frac{\sqrt{3(\Lambda_{4}-
	\frac{1}{2}\Lambda_{5})}}{4\pi G_{5}}
	+\frac{\omega}{a^{4}\sqrt{\Lambda_{4}-
	\frac{1}{2}\Lambda_{5}}}  ~,
\end{equation}
where the familiar Randall-Sundrum positive surface tension
term is accompanied now by a less familiar, but quite
characteristic, Regge-Teitelboim radiation-like term.
It has been already demonstrated\cite{dSbrane}, within
the framework of Regge-Teitelboim theory, that such a
radiation-like term can in fact be of a dynamical origin.
This calls, however, for a minimally coupled scalar field
serendipitously accompanied by a quartic potential.

\medskip
Several remarks are in order:

\medskip
\noindent $\bullet$
For the records, the Regge-Teitelboim dark radiation
term\cite{RTDavidson} was actually introduced before the
Randall-Sundrum dark radiation term\cite{RSdark}.
In spite of the similarity, however, the Regge-Teitelboim
dark radiation does not seem to have anything to do
with the Randall-Sundrum dark radiation.
The latter enters the theory once the FRW brane is
embedded\cite{RSdark} within (say) a Schwarzschild-AdS$_5$
black hole background
geometry.

\medskip
\noindent $\bullet$
Keep in mind that $\omega$ is not necessarily a small quantity.
Given the fact that $|\omega|$ parametrizes the deviation not
only from Randall-Sundrum model but from General Relativity
as well, its size will hopefully be fixed once the dark matter
interpretation discussed earlier is fully established.

\medskip
\noindent $\bullet$
At this stage, the sign of $\omega$ can still be either positive
or negative.
In the case of a maximally symmetric brane, for example, one
observes that  $\omega\rightarrow -\omega$ gives rise to
$\rho_{d}^{\displaystyle\ast}
\rightarrow-\rho_{d}^{\displaystyle\ast}$.
More general, however, as far as the FRW cosmic evolution is
concerned, one cannot really tell a theory associated with the pair
$\left\{\rho^{\displaystyle\ast},
\rho_{d}^{\displaystyle\ast}\right\}$ from its dual based on
$\left\{\rho^{\displaystyle\ast}+2\rho_{d}^{\displaystyle\ast},
-\rho_{d}^{\displaystyle\ast}\right\}$.

\medskip
The flat Minkowski brane is of special interest.
It is associated of course with setting $k=0$, and switching off
$\Lambda_4$, but in unified brane gravity, it also requires
starting from
\begin{equation}
	\rho (a)=\rho_{0}+
	 \sqrt{\frac{2}{-\Lambda_{5}}}
	 \frac{\omega}{a^{4}} ~,
	 \label{rhoflat}
\end{equation}
where $\rho_{0}$ stands for the fine-tuned (to make
$\Lambda_{4}$ vanish) Randall-Sundrum surface tension
\begin{equation}
	\rho_{0}=\frac{\sqrt{-6\Lambda_{5}}}{8\pi G_{5}}~.
\end{equation}
Note that, whereas Randall-Sundrum fine-tuning of the
brane tension is done at the Lagrangian level, the fine-tuning
of the additional radiation-like term depends on the value
of the would be constant of integration $\omega$.

\medskip
To lift the radiation fine-tuning, consider next small energy
density perturbations around a flat background, namely
\begin{equation}
	\rho =\left(\rho_{0}+
	\sqrt{\frac{2}{-\Lambda_{5}}}
	\frac{\omega}{a^{4}}\right)+
	\widetilde{\rho}~.
\end{equation}
This clearly includes the special case
$\displaystyle{\widetilde{\rho}=
\sqrt{\frac{2}{-\Lambda_{5}}}\frac{\Delta\omega}{a^4}}$,
but our interest in small $\widetilde{\rho}$ goes beyond
this case.
Let us thus define the small quantity $\epsilon$ 
\begin{equation}
	\epsilon=3 \frac{\dot{a}^{2}+k}{a^{2}}=8\pi G_{4}
	\left(\rho^{\displaystyle\ast}+
	\rho^{\displaystyle\ast}_{d} \right) ~,
	\label{epsilon}
\end{equation}
so that, following Eq.(\ref{omega}), 
\begin{equation}
	\begin{array}{l}
	\displaystyle{\rho^{\displaystyle\ast}_{d} \simeq
	-\sqrt{\frac{2}{-\Lambda_{5}}}\frac{\omega}{a^{4}}
	\left(1+\frac{\epsilon}{\Lambda_{5}}\right)}
	\vspace{6pt} \\
	\displaystyle{\rho^{\displaystyle\ast} \simeq
	\sqrt{\frac{2}{-\Lambda_{5}}}\frac{\omega}{a^{4}}
	\left(1+\frac{\epsilon}{\Lambda_{5}}\right)+
	\frac{\epsilon}{8\pi G_{4}}}~.
	\end{array}
\end{equation}
This is then substituted back into Eq.(\ref{stardef}), and
confronted with Eq.(\ref{epsilon}), and after some algebra
one arrives at
\begin{equation}
	\widetilde{\rho} =
	\left(\frac{1}{8\pi G_{4}}+\sqrt{\frac{6}{-\Lambda_{5}}}
	\left(\frac{1}{8\pi G_{5}}+
	\frac{\omega}{\sqrt{3}\Lambda_{5}a^{4}}\right)
	\right)\epsilon~.
\end{equation}
The fact that the coefficient of $\epsilon$ is not a constant
means that we are away from General Relativity.
It is only for large enough scale factors, namely for
\begin{equation}
	a(t) \gg \left(\frac{G_{5}\omega}
	{\Lambda_{5}}\right)^{1/4} ~,
\end{equation}
that one faces the low-energy Randall-Sundrum limit, and the
consequent identification Eq.(\ref{GN}) of the Newton constant.
It has not escaped our attention that for tiny $a(t)$, much
smaller than the above scale, a surplus amount
$\Delta \omega$ of radiation would naturally lead to an
effective brane cosmological constant $\Lambda^{eff}_{4}$
of order
\begin{equation}
	\Lambda^{eff}_{4}=
	{\cal O}\left(\frac{\Delta\omega}
	{\omega}\Lambda_{5}\right)~.
\end{equation}
The question whether such a 'inflation from radiation' scenario
can be fully matured, within the framework of a unified brane
model, into a satisfactory theory of inflation is still open.

\subsection{Randall-Sundrum limit}
The Randall-Sundrum limit is automatically built-in in
our theory.
To study the Regge-Teitelboim deviation from the
original Randall-Sundrum model, let us first solve
Eq.(\ref{cubic}) for $G_4\rightarrow\infty$.
In which case, we find
\begin{equation}
	\xi=\frac{2\pi G_{5}}{3}
	\left(\rho \pm \sqrt{\rho^2-
	\frac{\sqrt{3}\omega}{\pi G_{5}a^4}}\right) ~.
\end{equation}
The sign ambiguity of the quadratic equation can
be removed by insisting on reproducing the
Randall-Sundrum behavior at the $\omega\rightarrow 0$
limit.
Choosing the tenable plus sign, we find
\begin{equation}
	\frac{\dot{a}^2+k}{a^2}=
	\frac{\Lambda_5}{6}+
	\left(\frac{2\pi G_5}{3}\right)^2
	\left(\rho+\sqrt{\rho^2-
	\frac{\sqrt{3}\omega}{\pi G_{5} a^4}}\right)^2
	\label{RSomega}
\end{equation}
For small-$\omega$, $\xi$ can be approximated by
\begin{equation}
	\xi \simeq \frac{4\pi G_{5}}{3}\rho -
	\frac{\omega}{\sqrt{3}\rho a^{4}}~,
\end{equation}
so that the Randall-Sundrum expansion starts with
\begin{equation}
	\frac{\dot{a}^2+k}{a^2}\simeq
	\frac{\Lambda_5}{6}+
	\left(\frac{4\pi G_5}{3}\right)^2
	\rho^2-\frac{8\pi G_{5}\omega }{3\sqrt{3}a^4} 
	\label{RSlimit}
\end{equation}
In this limit, the effective energy density exhibits a so-called
dark
radiation term which highly reminds us of Randall-Sundrum
cosmology in a radially symmetric Schwarzschild-AdS
background.
The corresponding 5-dimensional Schwarzschild mass
being
\begin{equation}
	m=-\frac{8\pi G_{5}}{3\sqrt{3}}\omega ~.
\end{equation}
However, one should keep in mind that the two sources
of dark radiation are in fact independent of each other,
and once our analysis is extended beyond $Z_2$ symmetry,
both effects are expected to enter the game.

\medskip
Once $G_4$ turns finite, and this is the generic case,
Eq.(\ref{cubic}) becomes cubic.
Some algebra is then needed in order to single out the
particular solution which is the analytic continuation of
Eq.(\ref{RSlimit}).
Using the very convenient notation
\begin{equation}
	\chi (\rho)\equiv
	\sqrt{1-\frac{8\pi G^{2}_{5}}{3G_{4}}
	\left(\frac{\Lambda_{5}}{16\pi G_{4}}-\rho
	\right)} ~,
\end{equation}
assuming that $\rho$ is above some critical value (as
otherwise another solution takes over), the singled out
$\xi$ is approximated by	
\begin{equation}
	\xi\simeq\frac{G_{4}}{G_{5}}
	(\chi-1)-\frac{4\pi G^{2}_{5}\omega}{3\sqrt{3}G_{4}
	\chi(\chi-1)a^{4}}~,
\end{equation}
so that the small-$\omega$ Collins-Holdom expansion starts
now with
\begin{equation}
	\frac{\dot{a}^2+k}{a^2}\simeq
	\frac{\Lambda_5}{6}+
	\frac{G^{2}_{4}}{G^{2}_{5}} (\chi-1)^{2}-
	\frac{8\pi G_{5}\omega}{3\sqrt{3}\chi a^{4}} ~.
\end{equation}
In particular, for large enough $\rho(a)$, with the focus on the
early Universe, not only is Einstein gravity met again (owing to
$\chi\sim \sqrt{\rho}$), but the dark $\omega$-companion can
be in fact very different from a simple dark radiation term.

\subsection{Regge-Teitelboim limit}
Whereas the original Regge-Teitelboim limit is associated
with letting $G_{5}\rightarrow\infty$ while keeping
$\Lambda_{5}=0$, the modified Regge-Teitelboim limit
still calls for $G_{5}\rightarrow\infty$, but the brane correctly
separates now two tenable $\Lambda_{5}<0$ regions.
Eq.(\ref{omega}), which is reduced now to 
\begin{equation}
	\rho_{d}^{2}
	\left(8\pi G_{4}\left({\rho}+
	{\rho_{d}}\right)-
	\frac{\Lambda_{5}}{2}\right)=
	\frac{\omega^{2}}{a^{8}} ~,
\end{equation}
can be used to directly assign a dark companion
$\rho_{d}(\rho)$ to any given bare energy density $\rho$.

\medskip
An apparently empty brane, characterized by $\rho =0$
leading to $\rho_{d} \neq 0$, is clearly a special yet a very
pedagogical case.
In which case, the non-vanishing dark energy density
component behaves like
\begin{equation}
	\rho_{d}(a)\simeq \left\{
	\begin{array}{ll}
		\displaystyle{\left(\frac{\omega^{2}}
		{8\pi G_{4}a^{8}}\right)^{1/3}}
		 & ~~  \displaystyle{\rho_{d}\gg
		 \frac{-\Lambda_{5}}{16\pi G_{4}}} ~,
		 \vspace{6pt} \\
		\displaystyle{\frac{\sqrt{2}\omega}{\sqrt{
		-\Lambda_{5}}a^4}}
	 	  & ~~  \displaystyle{\rho_{d}\ll
		  \frac{-\Lambda_{5}}{16\pi G_{4}}} ~.
	\end{array}
	\right.
\end{equation}
The natural scale emerging here is
\begin{equation}
	a^{4}_{c}=\frac{16\sqrt{2}\pi G_{4}\omega}{(
	-\Lambda_{5}) ^{3/2}} ~.
\end{equation}
At very early times ($a\ll a_{c}$), the evolution is effectively
governed by a mysterious $P_{eff}=-\frac{1}{9}\rho_{eff}$
negative pressure dark cosmic background.
At late times ($a\gg a_{c}$), on the other hand, it is the
dark radiation which takes over.
The inclusion of a brane cosmological constant in this
game is straight forwards, and is easily achieved by shifting
$\Lambda_{5}\rightarrow \Lambda_{5}-2\Lambda_{4}$.
One can then argue that the transition from a
$P_{eff}=-\frac{1}{9}\rho_{eff}$ dominated Universe into
a $\Lambda_{4}$ dominated Universe (accompanied by a
leftover dark radiation of course) averagely resembles, at
least at its early stages, an effective dark matter era, meaning
$\rho_{d}\sim a^{-n}$ with $n\simeq 3$.
Such a unified dark matter/energy\cite{unified} idea has
already been demonstrated for the original
($\Lambda_{5}=0,~G_{5}\rightarrow\infty$) Regge-Teitelboim
model.

\section{Summary and conclusions}
The Israel Junction Conditions are known to be the major
theoretical tool in dealing with gravitational layers, shells, domain
walls, branes, etc.
In this paper, based on a fundamental brane variation principle,
as ingeniously prescribed by Dirac, we have actually challenged
the tight structure of the traditional gravitation matching equations.
Our main conclusion is that only a relaxed version of Israel
Junction Conditions is dictated in fact by the underlying constrained
gravity action.
The Israel junction conditions have been modified in analogy to
the Regge-Teitelboim modification of Einstein field equations.
To be a bit more technical, the tensorial Eq.(\ref{lambdaold}) has
been traded for the scalar $\cal K$-contracted Eq.(\ref{lambdanew})
and the conservation law Eq.(\ref{conserve}).            
It is important to emphasize that we do not claim, or even suggest,
that Israel matching equations are wrong; On the contrary, each
and every solution of Israel matching equations is strictly respected.
But following the present work, each such solution becomes a
representative of a wider continuous family of solutions.

\medskip
We have shown that the Dirac prescription for brane variation is
as vital as ever when bulk/brane gravity is switched on, and if
adopted, it may widely open the scope of modern brane theories.
The prototype example is provided by the Randall-Sundrum theory
which gets generalized accordingly.
The more so, Regge-Teitelboim and Randall-Sundrum brane
theories, which generalize General Relativity in two very different
theoretical directions, appear in fact to be two faces of the one
and the same unified brane theory.
Exercising the option of including a brane curvature term in the
Lagrangian, the unification scheme widen to also exhibit the
Collins-Holdom and Dvali-Gabadadze-Porrati limits.

\medskip
To appreciate the new formalism, the special case of a
$Z_{2}$-symmetric unified brane cosmology has been studied
in some details.
Here, the central equation is the cubic Eq.(\ref{cubic}) which allows
us to conveniently navigate within the
$\left( G_{4}^{-1},G_{5}^{-1},\omega \right)$ parameter space, with
all special cases easily accessible.
Not less important is Eq.(\ref{darkFRW}), which allows a physicist,
ignorant of the underlying unified brane theory, to recast the new
equations into Einstein's field equations format, and thus interpret
all deviations from General Relativity as a dark component.
It remains to be seen if this has (or unfortunately does not have)
anything to do with the real life dark matter problem.
Other interesting sub-topics are currently under extensive
investigation.
In particular, some novel insight on the inflationary era, carrying
the fingerprints of brane unification, will soon be released.
And the same holds of course, naturally with high expectations, for
the static radially symmetric case.

\acknowledgments{It is our pleasure to cordially thank Professors
Eduardo Guendelman and David Owen for enlightening discussions
and helpful comments. Special thanks to Shimon Rubin for sharing
with us the 'ups and downs' moments of this research, and
furthermore, for taking an active part in discussing several issues.}



\end{document}